\documentclass[lettersize,journal]{IEEEtran}
\usepackage{amsmath,amsfonts}
\usepackage{algorithmic}
\usepackage{algorithm}
\usepackage{array}
\usepackage{svg}
\usepackage[caption=false,font=normalsize,labelfont=sf,textfont=sf]{subfig}
\usepackage{textcomp}
\usepackage{stfloats}
\usepackage{url}
\usepackage{verbatim}
\usepackage{graphicx}
\usepackage{cite}
\usepackage{hyperref}
\usepackage{titlesec}
\usepackage{caption}
\usepackage{footmisc}

\usepackage{footnote}

\usepackage{tabularx,booktabs}
\newcolumntype{C}{>{\centering\arraybackslash}X} 

\titleformat*{\section}{\Large\bfseries}
\titleformat*{\subsection}{\large\bfseries}
\titleformat*{\subsubsection}{\large\bfseries}
\begin{document}

\title{Mammograms Classification: A Review}

\author{
    \IEEEauthorblockN{Marawan Elbatel\IEEEauthorrefmark{1}}
    
    \IEEEauthorblockA{\IEEEauthorrefmark{1}
    MSc in Medical Imaging and Applications, University of Girona.\\}
    
\thanks{This review paper was produced as part of Medical Sensors Project in the academic year 2021/2022 at the University of Burgundy, email(marwankefah@gmail.com)}
}



\maketitle

\begin{abstract}
An advanced reliable low-cost form of screening method, Digital mammography has been used as an effective imaging method for breast cancer detection. With an increased focus on technologies to aid healthcare, Mammogram images have been utilized in developing computer-aided diagnosis systems that will potentially help in clinical diagnosis. Researchers have proved that artificial intelligence with its emerging technologies can be used in the early detection of the disease and improve radiologists' performance in assessing breast cancer. In this paper, we review the methods developed for mammogram mass classification in two categories. The first one is classifying manually provided cropped region of interests (ROI) as either malignant or benign, and the second one is the classification of automatically segmented ROIs as either malignant or benign. We also provide an overview of datasets and evaluation metrics used in the classification task. Finally, we compare and discuss the deep learning approach to classical image processing and learning approach in this domain.
\end{abstract}
\section{\textbf{Introduction}}
\IEEEPARstart{H}IGHEST funded cancer in research in the US with low growth and high mortality rate, Breast cancer is seen by researchers as a key to understanding and researching the evolution of tumor cells \cite{Bray2018}. Therefore, early detection of this genetic disease will aid in further tumor analysis. In addition, having one woman diagnosed with breast cancer every 10 minutes, countries like the UK needs to raise attention to the treatment of the disease promptly. Breast Cancer, the most common cancer in women globally, has challenged researchers to find a treatment for it. With early detection as a major key in its treatment, Breast Cancer can be diagnosed in various ways, biopsy, ultrasound, mammograms, and MRI findings Early detection of the disease is the way to decrease the mortality rate. Breast cancer mortality rate is usually higher in underdeveloped and developing countries, where women are unaware and unable to access healthcare services. Breast cancer type whether benign or malignant depends on the cell that turns into cancer in the breast. For example, benign tumor cells have elliptical or round shapes with their edges being soft. On the other hand, malignant tumor cells tend to have spiculation, irregular shape, and ambiguous edges. Breast cancer is usually indicated by the existence of a mass with or without calcification accumulation. The radiologist usually examines the size of the mass, its shape, its location, its density to give a possible output about the nature of the mass whether it indicates cancer or not. Therefore, in a computer-aided diagnosis system, segmentation of the mass is essential for accurate and robust classification.
\subsection{Scope}
The number of papers released to aid the early detection of breast cancer in mammograms is incredibly high beyond the scope of the reasonably required paper length. Consequently, we limited our review to papers within high-impact factor journals as well as articles in the top journal's quartile (Q1). This survey provides a  review of classical image processing with machine learning and deep learning-based mammograms classification methods. The paper provides an in-depth analysis and explanation of the mammogram mass classification task in two categories. The first category is classifying regions of interest that are manually segmented as either malignant or benign. The second category by processing the whole mammogram for automated segmentation and classification of regions of interest as either malignant or benign. Both classical imaging processing and deep learning techniques are examined in each category.
\subsection{Outline of the Survey} \label{sec1}
The rest of this paper is organized as follows. In  \ref{problem} and \ref{challenges} , the problem of mammogram classification and its challenges are discussed. Datasets and evaluation metrics referenced in our chosen paper are listed in \ref{datasets}. In section \ref{manual}, we discuss the first category of the manually provided ROIs through both image processing and deep learning approaches.  Then, the second category of automated segmentation and classification is analyzed and reviewed in section \ref{automatic},  The paper is discussed and concluded  in section \ref{discussion} and  section \ref{Conclusion} respectively.

 \begin{table*}[!ht]
\begin{minipage}{\textwidth}

 \caption{Mammograms Datasets Comparison}
 \normalsize
\label{datasetsTable}
\begin{tabularx}{\textwidth}{@{}l*{7}{C}c@{}}

\toprule
Dataset &  Size & Different Resources & Description\footnote{The number of malignant findings and benign findings is reported for biopsy proven images}  \\ 
\midrule

NYU Dataset \cite{NYUdataset}{} &229,426 exams &No  &benign findings in 5,556 breast, malignant in 985, 236 with both. \\

DDSM \cite{Heath2000} &2,620 exams & Yes &benign findings in 96 exams, malignant in 89, 3 with both\\

MIAS \cite{suckling} &322 mammograms& Yes &benign findings in 61 images, malignant in 51, 1 with both\\

IRMA \cite{Oliveira2008} \footnote{We report the size of the dataset that is utlizied by the article reviewed} &2795 ROI & Yes &1863 abnormal ROI\\

\bottomrule
\end{tabularx}
\end{minipage}

\end{table*}

\section{Mammograms Breast Cancer Classification} \label{overview}
\subsection{The problem} \label{problem}
A low-dose X-ray examination, mammogram, is used for the early detection of breast cancer. Multiple studies demonstrated that mammograms decrease the mortality rate of breast cancer by providing an early diagnosis. However, performance benchmarks showed that nearly 80\% of biopsies performed after potential mammogram suspects are benign. Thus, the false-positive rate in mammograms is a critical problem that requires a more critical solution from researchers. The cost induced by the false positive rate is estimated to be several billion dollars. Thus, it is important to reduce the biopsy rates to reduce the costs as well as give more attention to real potential suspects. Moreover, mammogram reading by radiologists is tiring, requires a significant amount of time, is expensive, and is prone to errors. Double mammogram reading was found to improve the performance of classification, thus it can be proven that there is room for improvement in the classification of mammograms that a single reader did not reach yet. Consequently, work towards a computer-aided diagnosis system is essential for such a task.

\subsection{Main Challenges} \label{challenges}

The ideal mammogram classification task is to develop a general-purpose algorithm that achieves both optimal accuracy and robustness with high efficiency and a low false-positive rate. Classification efficiency depends on the segmentation quality of the region of interest in the mammogram. Deployment of the ideal algorithm in multiple settings requires the algorithm to be robust and potentially explainable.  With the rise of deep learning in the past few years and the high availability of data, Convolutional neural networks became the most popular method to be used in most computer vision problems. Breast cancer tumor cells come in different shapes and masses. Thus, the trade-off between accuracy and robustness of breast cancer models is also considered a major challenge and not yet well examined. Recent studies have shown that a model trained on a dataset fails to generalize when cross evaluated on another with small distributions shift. Comparative studies of models trained on different dataset distributions are not consistent and require researchers to develop a common benchmark for consistent comparison between models utilizing cross-evaluation to ensure that models do not exhibit data bias. Moreover, access to most of the datasets is difficult considering their confidentiality. Algorithms like federated learning have been released to train models to preserve the privacy of datasets \cite{Rieke2020}. However, it is prone to multiple privacy attacks. Imaging conditions vary also dependent on the device used to acquire those images. Such variations along with artifacts, poor resolution, and noise corruption can make it hard to reach the ideal general-purpose algorithm that will be able to generalize like a radiologist.

 \subsection{Datasets} \label{datasets}

 Most of the datasets for medical imaging usually suffer the problem of not being publicly available, thus making it hard to compare consistently model performances. Moreover, specific algorithms or techniques tend to show data-bias to specific dataset, making it hard for the algorithm to be applicable in other clinical settings. We examine here the datasets that are used and utilized in the reviewed articles as shown in Table \ref{datasetsTable}.A subset of the the NYU dataset is the NYU test set that also was used to form the NYU Reader Study Set. NYU Reader study set was mainly developed to compare model performance to that of radiologists. The study set contains labels collected from fourteen radiologist reader. Only the NYU dataset contains at least four images for each exam that corresponds to the four standard views in mammograms left and right craniocaudal, (L-CC), and (R-CC) respecitvely, and left and right mediolateral oblique ,(L-MLO) and (R-MLO) respectively.

\subsection{Performance Metrics}\label{metrics}

The metrics used in the reviewed articles are area under the receiver operating characteristic (AUC), areas under the precision-recall curve (AP), specificity, sensitivity, F1-score, and accuracy. Before exploring the metrics, we need to define True Positive $(TP)$, False Positive $(FP)$ ,True Negative $(TN)$and False Negative $(FN)$. 

$TP$ is the outcome when the algorithm classified the positive class correctly, a model output the mass is malignant and it is actually malignant.
\par
$FP$ is the outcome when the algorithm classify the negative class as being positive class, a model output the mass is malignant and it is actually not.
\par
$TN$ is the outcome when the algorithm classify the negative class correctly, a model output the mass is not malignant and it is not actually malignant.
\par

$FN$ is the outcome when the algorithm fail to detect the positive class correctly, a model output the mass is not malignant and it is actually malignant.

\par
\begin{equation}
Accuracy=\dfrac{TP+TN}{TP+TN+FP+FN} 
\label{recallEq}
\end{equation}
The specificity measures how good is your classifier in detecting the negative class.  
\begin{equation}
Specificity=\dfrac{TN}{TN+FP} 
\label{recallEq}
\end{equation}

\par
The sensitivity, recall, measures how good is your classifier in detecting the positive class.  
\begin{equation}
Sensitivity=\dfrac{TP}{TP+FN} 
\label{recallEq}
\end{equation}
\par
AUC is calculated as the Area Under the $Sensitivity$(TPR)-$(1-Specificity)$(FPR) Curve.

The F1-score is the harmonic mean of both the precision and recall described in equation \ref{f1Eq}.

\begin{equation}
F1=\dfrac{2*PR*Recall}{PR+Recall} 
\label{f1Eq}
\end{equation}

As the average precision calculated (AP) is the precision for 11 equally spaced recall values [0, 0.1, 0.2, 0.3 . . . 0.9, 1.0] on the Precision-Recall , a Pascal VOC metric described in equation \ref{apEq} and \ref{pIntrep} \cite{Everingham2009}.
\begin{equation}
AP_11=\dfrac{1}{11} \sum\limits_{R \in [0, 0.1,..., 1.0]} P_{intrep}(R)
\label{apEq}
\end{equation} 

\begin{equation}
P_{intrep}=max_{r':r'>r}(P(r'))
\label{pIntrep}
\end{equation}

\begin{figure*}[!t]
\centering
\captionsetup{justification=centering}
\subfloat[]{\includegraphics[width=5cm,height=3cm]{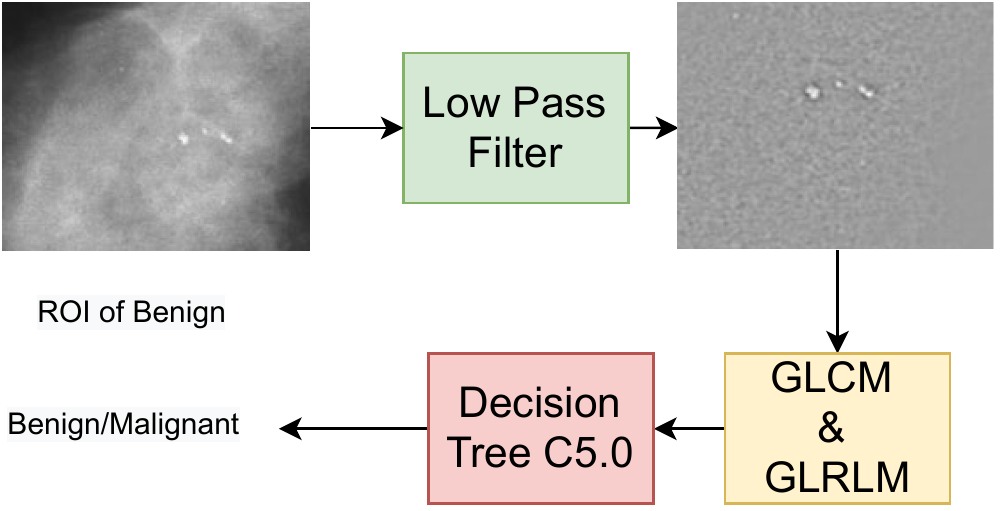}%
\label{mohantypipeline}}
\hfil
\hfil
\subfloat[]{\includegraphics[width=3cm,height=3cm]{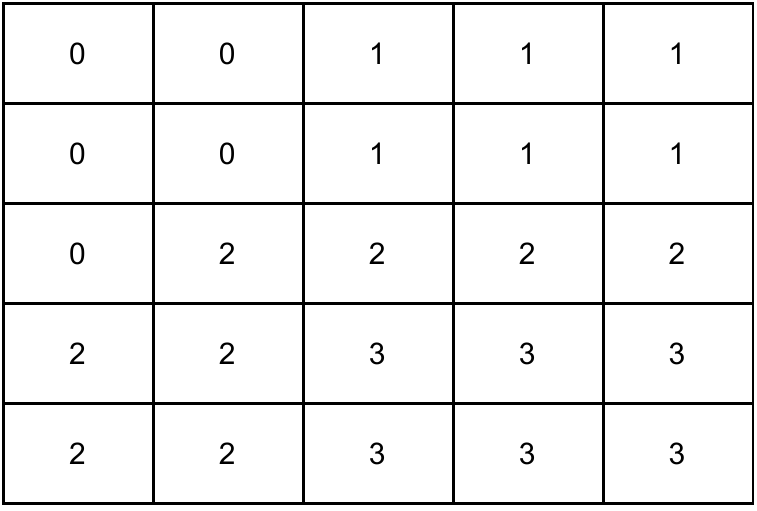}%
\label{matrixmoh}}
\hfil
\subfloat[]{\includegraphics[width=3cm,height=3cm]{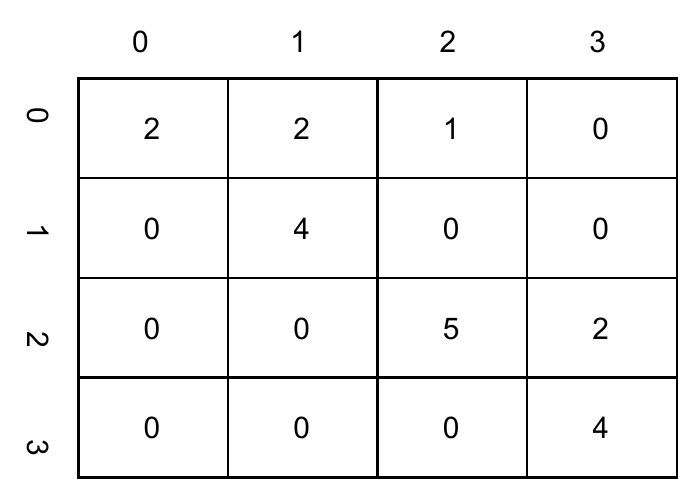}%
\label{glcm}}
\hfil
\subfloat[]{\includegraphics[width=3cm,height=3cm]{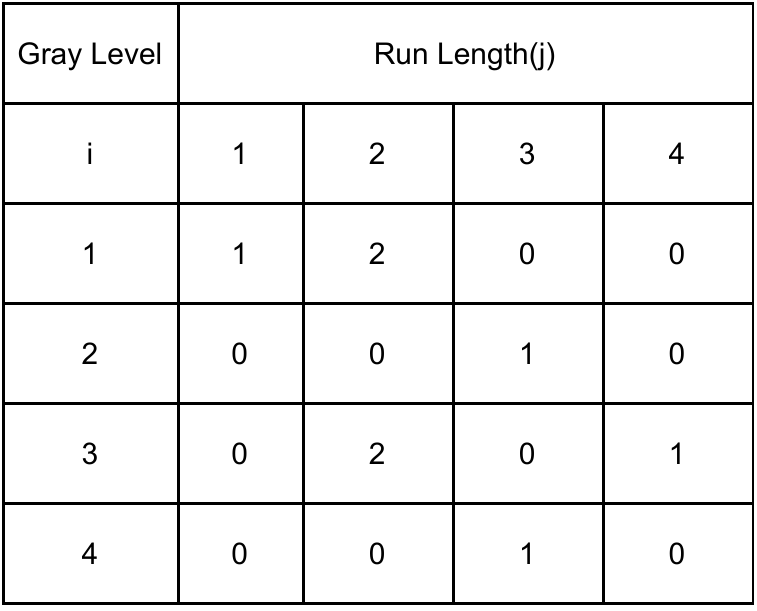}%
\label{glrl}}
\hfil
\subfloat[]{\includegraphics[width=10cm,height=3cm]{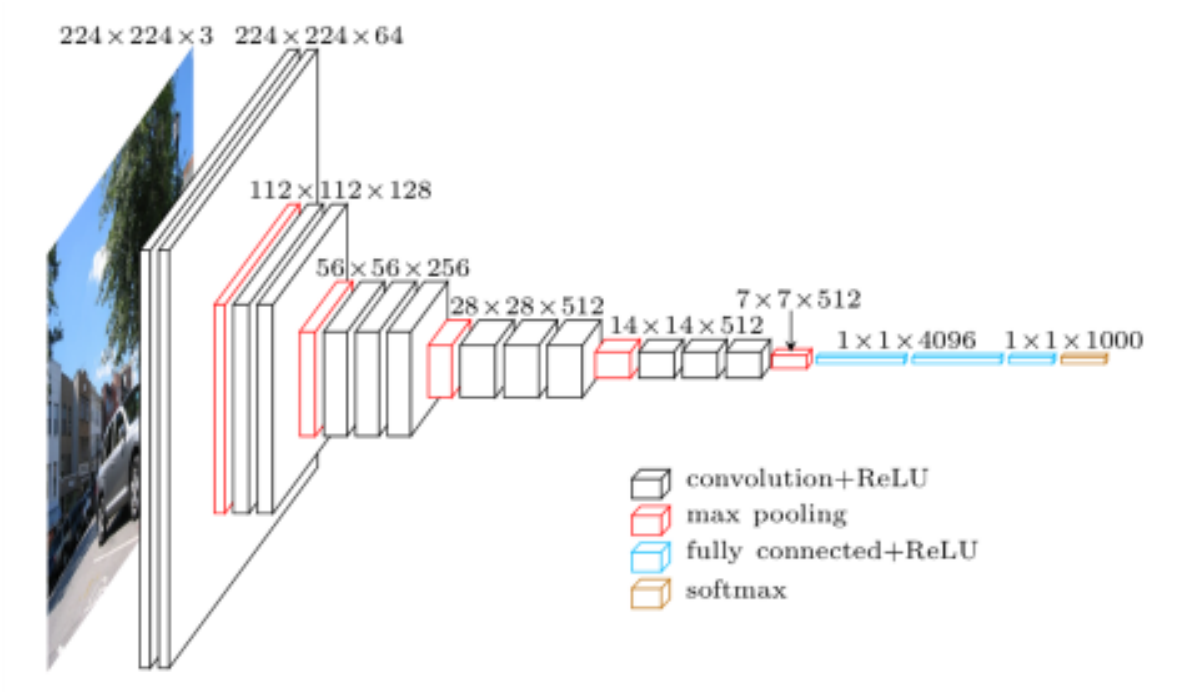}%
\label{vgg}}

\caption{(a) Mohanty et al Approach (b) 4 gray-level image example. (c) GLCM for
distance 1 and direction 0\textdegree. (d) GLRL matrix. e) Architecture for VGG-16 network.}

\end{figure*}

\section{Detection and classification of masses}\label{sec2}
Masses in mammograms are usually dense regions that differ in shape and size. A benign tumor mass usually has an elliptical or round shape with its edges being soft. Malignant tumors on the other hand tend to have spiculation, irregular shape, and ambiguous edges. Researchers have been developing algorithms to correctly detect and classify breast masses. We will present here the classification of masses after manual and automatic segmentation of regions of interest.

\subsection{Manually Cropped ROIs} \label{manual}
Mohanty et al reported on the use of texture-based features for the classification of regions of interest as either malignant or benign \cite{Mohanty2013}. Mohanty et al utilized the DDSM database for their classification purpose. The steps are taken to output their classification can be observed in figure \ref{mohantypipeline}. First, a low pass filter was used in the region of interest to suppress important features and preserve calcification within the mass. Then, a feature extraction approach was followed by computing the gray-level co-occurrence matrix (GLCM)
and gray-level run-length matrix (GLRLM) of the image. GLCM gives information about the texture of an image by generating a co-occurrence matrix that describes how often a combination of grey level values in an image appears in a specific direction within a specific distance as shown in figure \ref{glcm} \cite{Haralick1973}. Texture features can also be extracted from GLRLM that is a 2D matrix where each
entry in the matrix $X_ij$ is the number of elements j with the
grey level value i, in the direction j as shown in \ref{glrl} \cite{Tang1998}. 19 features were extracted from the two matrices. From the GLCM,14 features were extracted namely following Haralick et al intuition with correlation, entropy,
inertia, and homogeneity as the popular ones \cite{Haralick1973}. Another 5 texture features were extracted from GLRLM following Tang  suggestions \cite{Tang1998}. The texture is independent of the image intensity thus reflecting the intrinsic phenomenon of the image and making the feature extraction method robust to other data distributions. Texture features extracted from the images are then used to construct a C5.0 decision tree, which differs from the normal decision tree method by utilizing boosting, winnowing, and pruning in tree construction. Decision tree is based on the idea of constructing a conditional tree with test points as nodes with different branches. At each node, a decision is made to follow a specific branch and traverse down the tree. Finally, a leaf point is reached, and a prediction can be made. Their results can be shown on table \ref{resultstable} \cite{Pandya2015}.

Deep learning has succeeded in outperforming most classical imaging processing techniques in mammogram mass classification. Gardezi1 et al implemented a deep learning approach based on the VGG-16 CNN architecture to classify mammograms manually cropped ROIs as either normal or abnormal \cite{Gardezi2017}. Gardezi1 et al utilized the IRMA dataset in training and testing the model. VGG comes with different architecture but in general, it processes the image by a series of convolution and max-pooling layers to increase the receptor field and extract a feature vector that is usually an encoding of the input image \cite{Simonyan2015}. The used VGG-16 can be seen in figure \ref{vgg}. As the goal of the paper was to highlight the efficiency of deep learning models when used in the medical imaging domain, the model was not fine-tuned but original imageNet pre-trained weights were loaded. A feature matrix is then extracted at the first fully connected layer as shown in figure \ref{vgg}. The dimension of the feature matrix is 2795 × 4096, where 2795 is the total ROIs used, images batch, and 1x4096 as the encoding of each ROIs. Following, the feature matrix is then used to perform classification of the region as either normal or abnormal with four different classifiers, Support vector
machine (SVM), K nearest neighbor (KNN) classifier with k=1,3,5, binary decision tree, and simple logistic classifier. Summary of their important results for their methods is shown in Table. \ref{resultstable}

\subsection{Automatic segmentation and Classification}\label{automatic}
\begin{figure*}[!ht]
\centering
\captionsetup{justification=centering}
\subfloat[]{\includegraphics[width=5cm,height=5cm]{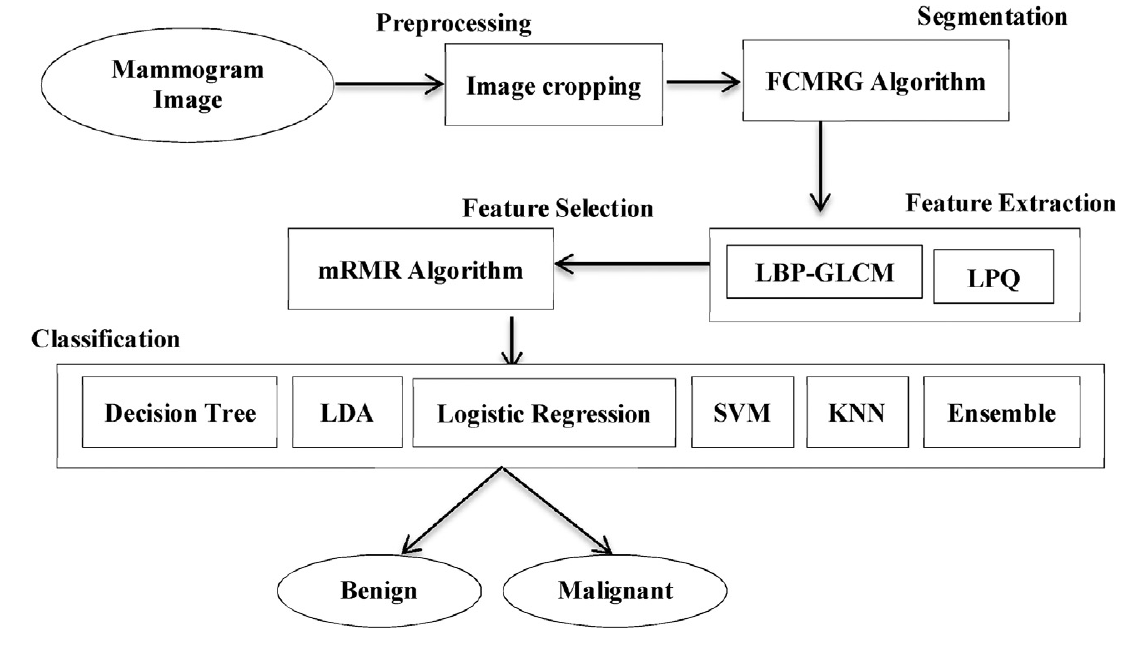}%
\label{sadadetal}}
\hfil
\subfloat[]{\includegraphics[width=4cm,height=6cm]{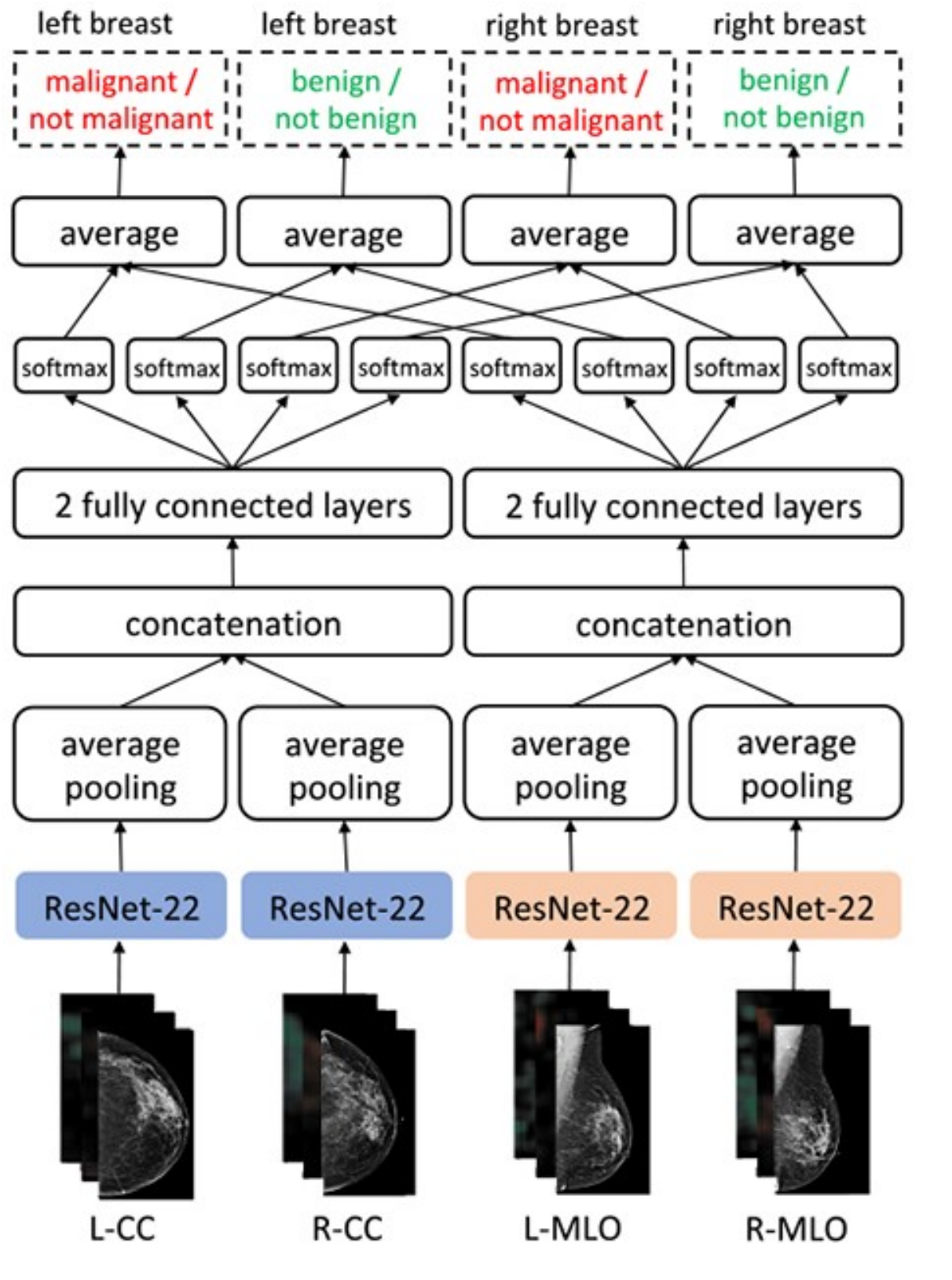}%
\label{DVCNN}}
\hfil
\subfloat[]{\includegraphics[width=7cm,height=6cm]{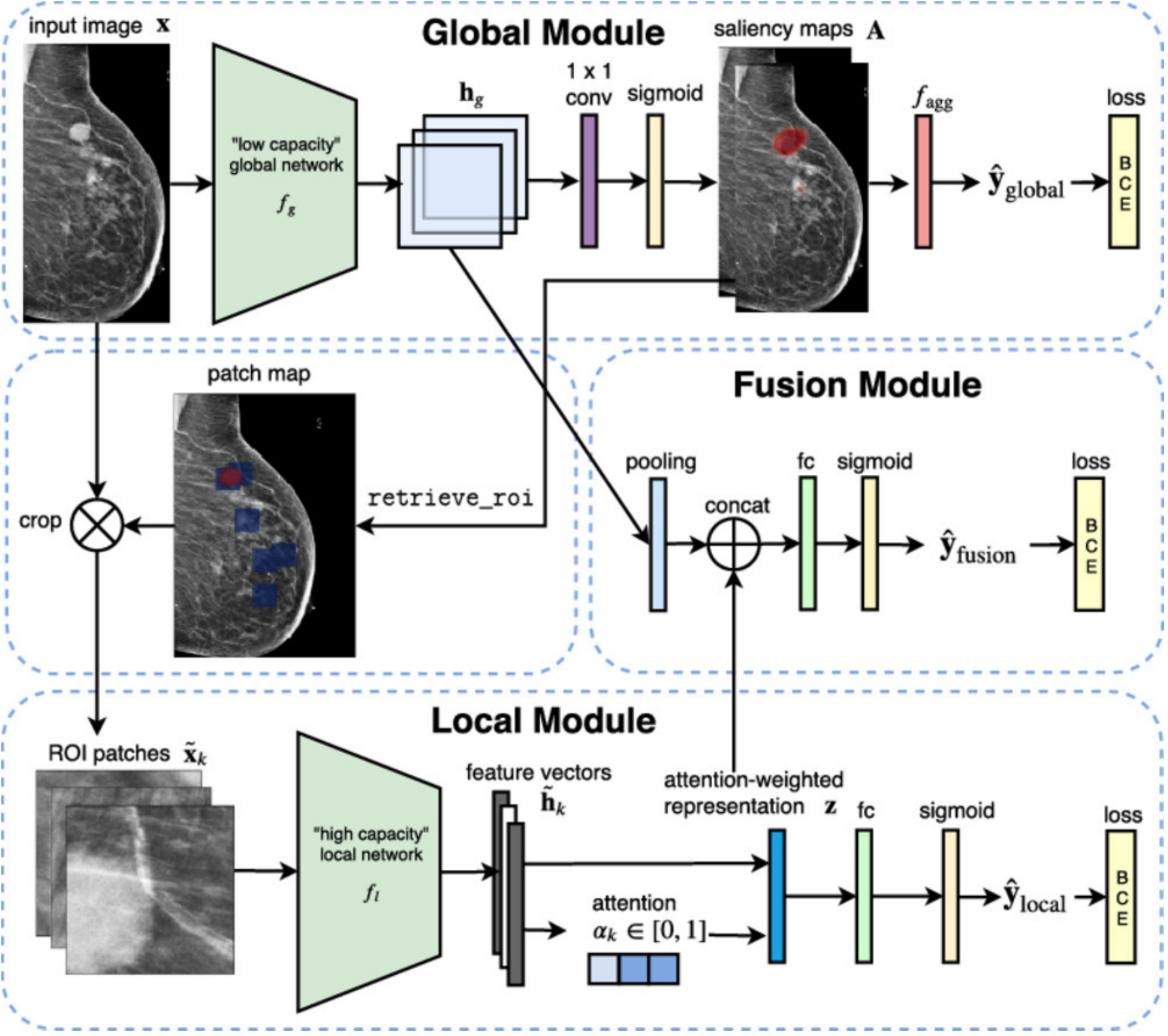}%
\label{gmic}}
\caption{(a) Sadada et al Proposed Method. (b) DV-CNN view model. (b) GMIC model architecture.}

\end{figure*}

Automated segmentation and classification of masses in mammograms is still being researched on a large scale. Sadada et al proposed a fuzzy c-means and region growing based automatic segmentation followed by hybrid texture feature extraction to aid in the classification of tumors as either benign or malignant \cite{Sadad2018}. Sadad et al pipeline can be shown in figure \ref{sadadetal}. The proposed method utilizes the two datasets DDSM and MIAS and was evaluated on 72 and 109 images from DDSM and MIAS respectively. The approach begins by firstly cropping the image to highlight the ROIs and suppress unwanted portions. Then, a newly described method called FCMRG was used. FCMRG refers to the use of fuzzy c means clustering followed by a region growing technique. Fuzzy c means clustering tends to assign membership to each pixel in every cluster \cite{Bezdek1984}. For example, An image with N pixels that is represented by set X =(x1, x2, . . ..., xN) is divided into L clusters. where every pixel has a membership value $r_ij$, where j is the pixel index and i is the clustered index. The goal is to minimize the loss function described in equation \ref{fuzzycmeansEq} where $ \sum\limits_{i=1}^{L} r_{ij}^{m}$. After iteratively updating the cluster centers and the membership value for every pixel as described in  until an accepted error margin for the loss function \cite{Bezdek1984}.
\begin{equation}
L_m= \sum\limits_{i=1}^{L} \sum\limits_{j=1}^{N} r_{ij}^{m} \mid x_j - c_i \mid ^2
\label{fuzzycmeansEq}
\end{equation}
 Cluster having the highest intensity values is then processed with a morphological operation to extract the tumor part. Then, a region-based algorithm is used to add the remaining pixels that were missed during extraction. After segmentation of the tumor part, two feature extraction methods followed, LBP-GLCM and LPQ. LBP-GLCM works by applying GLCM texture analysis on the image coded by the local binary patterns method instead of the original image. on the other hand, Heikkila et al introduced LPQ that depends on the characteristic of blur invariance of the Fourier phase spectrum, generating a code for every location forming a histogram similar to the LBP \cite{Ojansivu2008}. Texture features extracted from GLCM and LPQ are then fed to a Minimum-redundancy maximum-relevancy (mRMR)Algorithm proposed by that is used for feature selection \cite{Peng2005}. The algorithm obtains a subset of the feature vector with minimum redundancy and maximum relevance. The performance of classification with the selected features from LPQ. LBP-GLCM and hybrid features featuring both are evaluated and compared separately using four different classifiers, SVM, Logistic regression, linear discriminant analysis (LDA), decision trees, KNN, and an ensemble method. Summary of the results for the approach can be noted in Table \ref{resultstable}.

 \begin{table*}[h]
\begin{minipage}{\textwidth}
 \caption{Summary of Results of Reviewed Methods}
 \normalsize
\label{resultstable}
\begin{tabularx}{\textwidth}{ m{3cm}  m{2cm} m{3cm} m{3cm}  m{5cm}}

\toprule
Method &  Segmented ROIs &Output Classes & Dataset & Results \\ 
\midrule

GLCM \& GLRLM with DT \cite{Mohanty2013}{} & Manual &  Malignant/Benign& DDSM & 19 features reached 96.7\% accuracy with 0.995 AUC \\ \\

VGG-16 with different classifiers \cite{Gardezi2017}{} & Manual &  Normal/Abnormal & IRMA & 100 \% accuracy with 1 AUC for all classifiers slightly less for logisitc regression. \\ \\

Hybrid LBP-GLCM-LPQ with FCMRG segmentation \& different classifiers \cite{Sadad2018}{} & Manual &  Malignant/Benign & DDSM-MIAS & 98.2\% accuracy (0.98 AUC) for MIAS dataset with hybrid features by Decision Tree classifer and 93.1\% accuracy (0.97 AUC) for DDSM using SVM classifier  with only LPQ features.\\ \\

DMV-CNN \cite{Wu2020}{} & Automatic &  Malignant/Benign & NYU  Dataset & 0.779 AUC on NYU Reader Study, 0.824 AUC on NYU Test Set, and 0.532 AUC on DDSM. \\ \\

GMIC \cite{Shen2021}{} & Automatic &  Malignant/Benign & NYU Dataset & 0.857 AUC on NYU Reader Study, 0.901 AUC on NYU Test Set, and 0.604 AUC on DDSM. \\ \\

\bottomrule
\end{tabularx}
\end{minipage}
\end{table*}Deep learning on the other hand has been emerging in the classification of masses in mammograms in an incredible way. Wu et al revolutionized CAD systems by proposing a deep learning architecture that was proven as accurate as radiologists when evaluated on the same test set. Wu et al showed that an ensemble model utilizing radiologist prediction, as well as their model prediction, is more accurate than either separately \cite{Wu2020}. DMV-CNN proposed by Wu et al expect the four views of the exam to be present. The model was built after experimenting with four different model architectures with the proposed model concatenating L-CC and
R-CC representations, and L-MLO and R-MLO representations. The architecture of the model can be seen in figure \ref{DVCNN}. The model outputs prediction for MLO and CC separately then those predictions are averaged during inference. Thus, for each breast, the model output two probabilities, the probability of having a malignant tumor in that breast and the probability of having a benign tumor in that breast. The model was further enhanced by adding to the input channel of every view two heatmaps that were produced by
a patch-level classifier with DenseNet-121 architecture to get the benefit from pixel-level labels \cite{Huang2016}. Each of the two heatmaps corresponds to the probability of the pixel having a malignant tumor or benign tumor respectively. Surpassing
radiologists AUC by a margin of 0.11, Shen et al utilize a weakly supervised network in the classification of mammograms \cite{Shen2021}. Proposed by Shen et al, globally-aware multiple
instance classifier,(GMIC), works by extracting two saliency maps that correspond to the probability of having a benign or malignant tumor in that pixel from the high-resolution image with a global module that is a low capacity network then extract regions of interest from the saliency map to process with a local high capacity network, and in the end, makes a prediction based on the fusing of feature vectors from both branches. The low capacity network global network is implemented as ResNet-22 \cite{he2016deep}. The network was weakly supervised by training it only using image levels without pixel-level annotations. To map the saliency maps with the image-level labels to calculate the loss while training, a greedy aggregator function was used to efficiently map the output with the label and backpropagate the gradient efficiently. Saliency maps produced are used to retrieve regions of interest in the original high-resolution mammogram to form a patch of ROIs that is then fed to a high-capacity local module. The local module output feature vector for each ROI is then weighted using a gated attention network to be fused with feature maps from the global network to output the two final class predictions for the whole mammogram. The whole architecture of the model can be seen in figure \ref{gmic} and summary of the results is mentioned on Table \ref{resultstable}.

\section{Discussion} \label{discussion}
Medical images differs from the normal nature images in a variety of ways. For example, mammograms comes with  high resolution and the region of interests are very small compared to such resolution. Thus, many works that has been done for nature images may not be applicable in the medical imaging domain. Image processing with classical machine learning techniques tend to be effective in the classification of mammograms. However, deep learning outperformed all previous techniques as well as radiologists' classification in such task. Deep learning can surpass human level in vision task applications ,but the problems of deep learning remains unsolved till our day. Deep learning suffers from the problem of inexplainble output as it can not be divided to smaller explainable tasks as classical image processing techniques. Also, it required a high memory and computational power for usage in clincal settings.  Moreover, the robustness and generalization capabilities of deep learning has been an issue for years. A model trained on a specific distribution fails to generalize when cross evaluated on another. Mammograms is acquired from different populations and different machines, thus making it difficult to trust the deep learning models completely in such domain. The future trend works on gathering more data for training in a private decentralized manner, federated learning, towards a robust generalized model that can be deployed in more than one setting \cite{Rieke2020}.  
\section{Conclusion} \label{Conclusion}

The vast emerging research in the mammogram classification has proved that CAD systems will improve clinical diagnosis in a significant way in the future. image processing techniques with classical machine learning has been reviewed and discussed with the deep learning approach in the classification of mammograms. Deep learning has been also compared extensively with other approaches in our discussion.

\bibliography{ref}

\begin{thebibliography}{10}
\providecommand{\url}[1]{#1}
\csname url@samestyle\endcsname
\providecommand{\newblock}{\relax}
\providecommand{\bibinfo}[2]{#2}
\providecommand{\BIBentrySTDinterwordspacing}{\spaceskip=0pt\relax}
\providecommand{\BIBentryALTinterwordstretchfactor}{4}
\providecommand{\BIBentryALTinterwordspacing}{\spaceskip=\fontdimen2\font plus
\BIBentryALTinterwordstretchfactor\fontdimen3\font minus
  \fontdimen4\font\relax}
\providecommand{\BIBforeignlanguage}[2]{{%
\expandafter\ifx\csname l@#1\endcsname\relax
\typeout{** WARNING: IEEEtran.bst: No hyphenation pattern has been}%
\typeout{** loaded for the language `#1'. Using the pattern for}%
\typeout{** the default language instead.}%
\else
\language=\csname l@#1\endcsname
\fi
#2}}
\providecommand{\BIBdecl}{\relax}
\BIBdecl

\bibitem{Bray2018}
\BIBentryALTinterwordspacing
F.~Bray, J.~Ferlay, I.~Soerjomataram, R.~L. Siegel, L.~A. Torre, and A.~Jemal,
  ``Global cancer statistics 2018: Globocan estimates of incidence and
  mortality worldwide for 36 cancers in 185 countries,'' \emph{CA: a cancer
  journal for clinicians}, vol.~68, pp. 394--424, 11 2018. [Online]. Available:
  \url{https://pubmed.ncbi.nlm.nih.gov/30207593/}
\BIBentrySTDinterwordspacing

\bibitem{NYUdataset}
\BIBentryALTinterwordspacing
N.~Wu, J.~Phang, J.~Park, Y.~Shen, S.~G. Kim, L.~Heacock, L.~Moy, K.~Cho, and
  K.~J. Geras, ``The nyu breast cancer screening dataset v1.0,'' 2019.
  [Online]. Available: \url{https://cs.nyu.edu/~kgeras/reports/datav1.0.pdf.}
\BIBentrySTDinterwordspacing

\bibitem{Heath2000}
M.~D. Heath, K.~Bowyer, D.~Kopans, and R.~H. Moore, ``The digital database for
  screening mammography,'' \emph{Proceedings of the 5th International Workshop
  on Digital Mammography}, vol. Medical Physics Pubâ€¦, pp. 212--218, 2000.

\bibitem{suckling}
J.~Suckling, J.~Parker, D.~Dance, S.~Astley, I.~Hutt, C.~Boggis, I.~Ricketts,
  N.~Cerneaz, S.~Kok, P.~Taylor, D.~Betal, and J.~Savage, ``The mammographic
  image analysis society digital mammogram database,'' \emph{Exerpta Medica
  International Congress}, vol. 1069, pp. 375--378, 1994.

\bibitem{Oliveira2008}
J.~E.~E. Oliveira, M.~O. Gueld, A.~de~A.~AraÃºjo, B.~Ott, and T.~M. Deserno,
  ``Toward a standard reference database for computer-aided mammography,''
  \emph{https://doi.org/10.1117/12.770325}, vol. 6915, pp. 606--614, 3 2008.

\bibitem{Rieke2020}
\BIBentryALTinterwordspacing
N.~Rieke, J.~Hancox, W.~Li, F.~MilletarÃ¬, H.~R. Roth, S.~Albarqouni,
  S.~Bakas, M.~N. Galtier, B.~A. Landman, K.~Maier-Hein, S.~Ourselin,
  M.~Sheller, R.~M. Summers, A.~Trask, D.~Xu, M.~Baust, and M.~J. Cardoso,
  ``The future of digital health with federated learning,'' \emph{npj Digital
  Medicine 2020 3:1}, vol.~3, pp. 1--7, 9 2020. [Online]. Available:
  \url{https://www.nature.com/articles/s41746-020-00323-1}
\BIBentrySTDinterwordspacing

\bibitem{Everingham2009}
\BIBentryALTinterwordspacing
M.~Everingham, L.~V. Gool, C.~K.~I. Williams, J.~Winn, and A.~Zisserman, ``The
  pascal visual object classes (voc) challenge,'' \emph{International Journal
  of Computer Vision 2009 88:2}, vol.~88, pp. 303--338, 9 2009. [Online].
  Available: \url{https://link.springer.com/article/10.1007/s11263-009-0275-4}
\BIBentrySTDinterwordspacing

\bibitem{Mohanty2013}
A.~K. Mohanty, M.~R. Senapati, S.~Beberta, and S.~K. Lenka, ``Texture-based
  features for classification of mammograms using decision tree,'' \emph{Neural
  Computing and Applications}, vol.~23, pp. 1011--1017, 9 2013.

\bibitem{Haralick1973}
R.~M. Haralick, I.~Dinstein, and K.~Shanmugam, ``Textural features for image
  classification,'' \emph{IEEE Transactions on Systems, Man and Cybernetics},
  vol. SMC-3, pp. 610--621, 1973.

\bibitem{Tang1998}
X.~Tang, ``Texture information in run-length matrices,'' \emph{IEEE
  Transactions on Image Processing}, vol.~7, pp. 1602--1609, 1998.

\bibitem{Pandya2015}
R.~Pandya, J.~Pandya, K.~P. Dholakiya, and I.~Amreli, ``C5.0 algorithm to
  improved decision tree with feature selection and reduced error pruning,''
  \emph{International Journal of Computer Applications}, vol. 117, pp.
  975--8887, 2015.

\bibitem{Gardezi2017}
S.~J.~S. Gardezi, M.~Awais, I.~Faye, and F.~Meriaudeau, ``Mammogram
  classification using deep learning features,'' \emph{Proceedings of the 2017
  IEEE International Conference on Signal and Image Processing Applications,
  ICSIPA 2017}, pp. 485--488, 2017.

\bibitem{Simonyan2015}
\BIBentryALTinterwordspacing
K.~Simonyan, ``Very deep convolutional networks for large-scale image
  recognition.'' 2015. [Online]. Available:
  \url{http://arxiv.org/abs/1409.1556}
\BIBentrySTDinterwordspacing

\bibitem{Sadad2018}
T.~Sadad, A.~Munir, T.~Saba, and A.~Hussain, ``Fuzzy c-means and region growing
  based classification of tumor from mammograms using hybrid texture feature,''
  \emph{Journal of Computational Science}, vol.~29, pp. 34--45, 11 2018.

\bibitem{Bezdek1984}
J.~C. Bezdek, R.~Ehrlich, and W.~Full, ``Fcm: The fuzzy c-means clustering
  algorithm,'' \emph{Computers \& Geosciences}, vol.~10, pp. 191--203, 1 1984.

\bibitem{Ojansivu2008}
\BIBentryALTinterwordspacing
V.~Ojansivu and J.~HeikkilÃ¤, ``Blur insensitive texture classification using
  local phase quantization,'' \emph{Lecture Notes in Computer Science
  (including subseries Lecture Notes in Artificial Intelligence and Lecture
  Notes in Bioinformatics)}, vol. 5099 LNCS, pp. 236--243, 2008. [Online].
  Available:
  \url{https://link.springer.com/chapter/10.1007/978-3-540-69905-7_27}
\BIBentrySTDinterwordspacing

\bibitem{Peng2005}
H.~Peng, F.~Long, and C.~Ding, ``Feature selection based on mutual information:
  Criteria of max-dependency, max-relevance, and min-redundancy,'' \emph{IEEE
  Transactions on Pattern Analysis and Machine Intelligence}, vol.~27, pp.
  1226--1238, 8 2005.

\bibitem{Wu2020}
N.~Wu, J.~Phang, J.~Park, Y.~Shen, Z.~Huang, M.~Zorin, S.~Jastrzebski,
  T.~Fevry, J.~Katsnelson, E.~Kim, S.~Wolfson, U.~Parikh, S.~Gaddam, L.~L.~Y.
  Lin, K.~Ho, J.~D. Weinstein, B.~Reig, Y.~Gao, H.~Toth, K.~Pysarenko,
  A.~Lewin, J.~Lee, K.~Airola, E.~Mema, S.~Chung, E.~Hwang, N.~Samreen, S.~G.
  Kim, L.~Heacock, L.~Moy, K.~Cho, and K.~J. Geras, ``Deep neural networks
  improve radiologists' performance in breast cancer screening,'' \emph{IEEE
  Transactions on Medical Imaging}, vol.~39, pp. 1184--1194, 4 2020.

\bibitem{Shen2021}
Y.~Shen, N.~Wu, J.~Phang, J.~Park, K.~Liu, S.~Tyagi, L.~Heacock, S.~G. Kim,
  L.~Moy, K.~Cho, and K.~J. Geras, ``An interpretable classifier for
  high-resolution breast cancer screening images utilizing weakly supervised
  localization,'' \emph{Medical Image Analysis}, vol.~68, p. 101908, 2 2021.

\bibitem{Huang2016}
\BIBentryALTinterwordspacing
G.~Huang, Z.~Liu, L.~V.~D. Maaten, and K.~Q. Weinberger, ``Densely connected
  convolutional networks,'' \emph{Proceedings - 30th IEEE Conference on
  Computer Vision and Pattern Recognition, CVPR 2017}, vol. 2017-January, pp.
  2261--2269, 8 2016. [Online]. Available:
  \url{https://arxiv.org/abs/1608.06993v5}
\BIBentrySTDinterwordspacing

\bibitem{he2016deep}
K.~He, X.~Zhang, S.~Ren, and J.~Sun, ``Deep residual learning for image
  recognition,'' in \emph{Proceedings of the IEEE conference on computer vision
  and pattern recognition}, 2016, pp. 770--778.

\end{thebibliography}
\bibliographystyle{IEEEtran}


\vfill

\end{document}